# A note on the role of charge conservation in electronegativity equalization and its implications for the translational symmetries of electrostatic properties in fluctuating-charge models

Jiahao Chen[1] and Todd J. Martínez[1,2]

[1]Department of Chemistry and Frederick Seitz Materials Research Laboratory, University of Illinois at Urbana-Champaign, Urbana, IL

[2]Department of Chemistry, Stanford University, Stanford, CA and SLAC National Accelerator Laboratory, Menlo Park, CA

## 1 Introduction

Fluctuating-charge models are a computationally efficient way to treat polarization and charge transfer effects in molecular mechanics,[1-6] the omission of which can lead to qualitative errors in dynamical simulations.[7-10] This is accomplished by applying the principle of electronegativity equalization[11,12] to recalculate an entire system's charge distribution in response to nuclear configuration changes at every time step. It is well understood that the presence of different atomic electronegativities plays an integral role in electronegativity equalization; however, the equally important constraint of charge conservation, while always accounted for in calculating the charge distribution numerically, has not always been well appreciated. In fact, charge conservation is critical for providing the correct spatial transformation properties of dipole moments and polarizabilities.

In this Note, we present a short derivation of an analytic solution that distinguishes the role of charge conservation from the much better appreciated effects of electronegativity differences in the process of electronegativity equalization. An analytic solution has previously been published for the ES+ model, albeit without derivation;[13] however, its physical significance in representing these two distinct effects remains underappreciated. This has significant repercussions for the calculation of electrostatic properties in the context of fluctuating charge models, and there has been significant confusion over this topic in the literature. In particular, many published formulae for dipole moments and polarizabilities do not exhibit the correct translational symmetries without special choices of coordinate origin, and it is often necessary to select the origin carefully to avoid spurious coordinate dependence of these electrostatic properties.[11,12,20-22] It is difficult to reconcile such choices with the translational and rotational symmetries required by classical electrostatics, which require that polarizabilities and dipole moments (for neutral systems, in the latter case) be translationally invariant.[25,26] As it turns out, the correct treatment of the terms arising from charge conservation solves this problem naturally.

In this Note, we show how the method of Lagrange multipliers and Gaussian elimination produce an analytic solution for the charge distribution that contains two different terms, clearly separating the contribution of charge conservation from that of charge–charge interactions and chemical hardness. This allows us to identify the roles of these separate terms in electrostatic observables such as dipole moments and polarizabilities.

## 2 Analytic solution of fluctuating–charge models

A fluctuating-charge model is solved by a charge distribution that minimizes an energy expression. In most modern models, this energy is quadratic in the charges and takes the form

$$E(\mathbf{q}) = \sum_{i=1}^{N} q_i \chi_i + \frac{1}{2} \sum_{i,j=1}^{N} q_i q_j \eta_{ij} = \mathbf{q}^T \boldsymbol{\chi} + \frac{1}{2} \mathbf{q}^T \boldsymbol{\eta} \mathbf{q} \qquad [1]$$

where $q_i$ is the charge on atom $i$, $\chi_i$ is the intrinsic Mulliken electronegativity[14-16] of atom $i$, $\eta_{ii}$ is the chemical hardness[17] of atom $i$, and $\eta_{ij}$ are screened Coulomb interactions, the details of which vary from model to model.[3] We also introduce the boldface convention for vectors and matrices acting in the space of atomic charge variables, i.e. $\mathbf{q}$ and $\boldsymbol{\chi}$ are $N$–vectors and $\boldsymbol{\eta}$ is a $N \times N$ real and symmetric matrix. We have previously noted[18] that there is an exact analogy in classical electric circuits, where atoms can be interpreted as serial pairs of batteries of voltage $\chi_i$ and capacitors of capacitance $\eta_{ii}^{-1}$, which are coupled with coefficients of inductance $\eta_{ij}^{-1}$. This expression is formally equivalent to a Taylor series expansion of the energy with respect to the atomic charges to quadratic order.[19] Thus assuming that the hardness matrix $\boldsymbol{\eta}$ is invertible, and after discarding an irrelevant constant term, the energy can be expressed as the quadratic form

$$E(\mathbf{q}) = \frac{1}{2} \left( \mathbf{q} + \boldsymbol{\eta}^{-1} \boldsymbol{\chi} \right)^T \boldsymbol{\eta} \left( \mathbf{q} + \boldsymbol{\eta}^{-1} \boldsymbol{\chi} \right) \qquad [2]$$

At first blush, it is tempting to note that this quadratic form is minimized by the solution

$$\mathbf{q} = \mathbf{q}_u = -\boldsymbol{\eta}^{-1} \boldsymbol{\chi} \qquad [3]$$

and hence assert that $\mathbf{q}_u$ is the solution to the fluctuating-charge model. However, this solution does not account for charge conservation, which is essential for electronegativity equalization.[11,12,20-22] (Thus, we have used the subscript u to denote an unconstrained solution.) This physical conservation law imposes a constraint on the total charge of the system

$$\sum_{i=1}^{N} q_i = \mathbf{q}^T \mathbf{1} = Q \qquad [4]$$

where $Q$ is the total charge and $\mathbf{1}$ is a column $N$–vector with all entries equal to unity. We use the method of Lagrange multipliers to reformulate the original problem as an unconstrained minimization:[23] by introducing the Lagrange multiplier $\mu$, which has a physical interpretation as the chemical potential of charge, we construct and minimize the Lagrange function

$$F(\mathbf{q},\mu) = E(\mathbf{q}) - \mu(\mathbf{q}^T \mathbf{1} - Q) = \mathbf{q}^T(\boldsymbol{\chi} - \mu\mathbf{1}) + \frac{1}{2}\mathbf{q}^T \boldsymbol{\eta} \mathbf{q} + \mu Q \qquad [5]$$

Minimization of this Lagrange function leads to a linear system which can be written in the $2 \times 2$ block matrix form

$$\begin{pmatrix} \boldsymbol{\eta} & \mathbf{1} \\ \mathbf{1}^T & 0 \end{pmatrix} \begin{pmatrix} \mathbf{q} \\ \mu \end{pmatrix} = \begin{pmatrix} -\boldsymbol{\chi} \\ Q \end{pmatrix} \qquad [6]$$

where again we note that $\boldsymbol{\eta}$ is a $N \times N$ matrix; $\mathbf{1}, \mathbf{q}$ and $\boldsymbol{\chi}$ are column $N$–vectors, and $0, \mu$ and $Q$ are scalars. Thus we have converted a constrained optimization in $N$ unknowns to a linear system of $N + 1$ unknowns; this type of linear system is known in the numerical analysis literature as a saddle point problem, and many computationally efficient methods have been developed for solving such problems numerically.[24] We now use Gaussian elimination to derive an analytic solution. As before, we assume that the hardness matrix $\boldsymbol{\eta}$ is invertible, pre-multiply the first row by $-\mathbf{1}^T \boldsymbol{\eta}^{-1}$ and add the resulting equation to the second row. After some rearrangement, we obtain the solution

$$\begin{pmatrix} \mathbf{q} \\ \mu \end{pmatrix} = \begin{pmatrix} -\boldsymbol{\eta}^{-1}(\boldsymbol{\chi} + \mu \mathbf{1}) \\ -\dfrac{Q + \mathbf{1}^T \boldsymbol{\eta}^{-1} \boldsymbol{\chi}}{\mathbf{1}^T \boldsymbol{\eta}^{-1} \mathbf{1}} \end{pmatrix} \quad [7]$$

from which it is immediately clear that the solution $\mathbf{q}$ differs from the unconstrained solution $\mathbf{q}_u$ in Eq. [3] by an additional term $-\mu \boldsymbol{\eta}^{-1} \mathbf{1}$ that arises directly from the charge conservation constraint of Eq. [4]. Thus $-\boldsymbol{\eta}^{-1} \boldsymbol{\chi}$ represents the driving effect of electronegativity differences while the other term $-\mu \boldsymbol{\eta}^{-1} \mathbf{1}$ captures the restrictions imposed by charge conservation.

## 3  Analytic formulae for dipole moment and polarizabilities

We now study how the contributions of electronegativity differences and charge conservation contribute in the calculation of electrostatic properties such as multipole moments and polarizabilities. The dipole moment can be obtained immediately from the definition

$$d_\lambda = \sum_{i=1}^{N} q_i R_{i\lambda} = \mathbf{q}^T \mathbf{R}_\lambda \quad [8]$$

where the Greek index $\lambda$ denotes a spatial component and $R_{i\lambda}$ is the $\lambda^{\text{th}}$ spatial component of the position of atom $i$. To obtain the dipole polarizability, we use the method of finite fields and employ the usual dipole coupling prescription to construct the energy in the presence of an external electrostatic field $\vec{\varepsilon}$ as

$$E(\mathbf{q}; \varepsilon_\lambda) = E(\mathbf{q}) - \mathbf{q}^T \mathbf{R}_\lambda \varepsilon_\lambda = \mathbf{q}^T (\boldsymbol{\chi} - \mathbf{R}_\lambda \varepsilon_\lambda) + \frac{1}{2} \mathbf{q}^T \boldsymbol{\eta} \mathbf{q} \quad [9]$$

where we have used the Einstein implicit summation convention for repeated Greek indices. The external field simply perturbs the atomic electronegativities by an amount $\mathbf{R}_\lambda \varepsilon_\lambda$ which is the potential produced by the external field. Therefore, we can replace $\boldsymbol{\chi}$ by $\boldsymbol{\chi} - \mathbf{R}_\lambda \varepsilon_\lambda$ in [7] to obtain the new charge distribution as

$$\begin{pmatrix} \mathbf{q}(\varepsilon_\lambda) \\ \mu(\varepsilon_\lambda) \end{pmatrix} = \begin{pmatrix} -\boldsymbol{\eta}^{-1}(\boldsymbol{\chi} - \mathbf{R}_\lambda \varepsilon_\lambda + \mu(\varepsilon_\lambda)\mathbf{1}) \\ -\dfrac{Q + \mathbf{1}^T \boldsymbol{\eta}^{-1}(\boldsymbol{\chi} - \mathbf{R}_\lambda \varepsilon_\lambda)}{\mathbf{1}^T \boldsymbol{\eta}^{-1}\mathbf{1}} \end{pmatrix} \qquad [10]$$

which corresponds to an energy of

$$E_0(\varepsilon_\lambda) = -\frac{1}{2}(\boldsymbol{\chi} - \mathbf{R}_\lambda \varepsilon_\lambda)^T \boldsymbol{\eta}^{-1}(\boldsymbol{\chi} - \mathbf{R}_\lambda \varepsilon_\lambda) + \frac{1}{2}\frac{(Q + \mathbf{1}^T \boldsymbol{\eta}^{-1}(\boldsymbol{\chi} - \mathbf{R}_\lambda \varepsilon_\lambda))^2}{\mathbf{1}^T \boldsymbol{\eta}^{-1}\mathbf{1}} \qquad [11]$$

We can verify that the dipole moment [8] is the expected derivative of $E_0$ with respect to the external field,

$$d_\lambda = \left.\frac{\partial E_0}{\partial \varepsilon_\lambda}\right|_{\varepsilon_\lambda = 0} = \left[-\mathbf{R}_\lambda^T \boldsymbol{\eta}^{-1}(\boldsymbol{\chi} - \vec{\mathbf{R}} \cdot \vec{\varepsilon}) - \mu(\vec{\varepsilon})(\mathbf{1}^T \boldsymbol{\eta}^{-1}\mathbf{R}_\lambda)\right]_{\varepsilon_\lambda = 0} \qquad [12]$$
$$= -\mathbf{R}_\lambda^T \boldsymbol{\eta}^{-1}\boldsymbol{\chi} - \mu \mathbf{R}_\lambda^T \boldsymbol{\eta}^{-1}\mathbf{1}$$

The dipole polarizability is the next derivative,

$$\alpha_{\lambda\rho} = \left.\frac{\partial d_\rho}{\partial \varepsilon_\lambda}\right|_{\varepsilon_\lambda = 0} = \left.\frac{\partial E_0}{\partial \varepsilon_\lambda \partial \varepsilon_\rho}\right|_{\varepsilon_\lambda = 0} = -\mathbf{R}_\rho^T \boldsymbol{\eta}^{-1}\mathbf{R}_\lambda - \frac{(\mathbf{1}^T \boldsymbol{\eta}^{-1}\mathbf{R}_\lambda)(\mathbf{1}^T \boldsymbol{\eta}^{-1}\mathbf{R}_\rho)}{\mathbf{1}^T \boldsymbol{\eta}^{-1}\mathbf{1}} \qquad [13]$$

Again, these formulae differ from the corresponding ones based on $\mathbf{q}_u$ by an extra term arising from the charge constraint. When these terms are omitted, the dipole moment and dipole polarizability have incorrect translational symmetries; retaining these terms, in contrast, results in expressions that exhibit the correct symmetries. Indeed, we see that under the global coordinate translation $\vec{\mathbf{R}} \mapsto \vec{\mathbf{R}} - \vec{s}\mathbf{1}$, the dipole moment transforms as

$$d_\lambda \mapsto d_\lambda + s_\lambda \mathbf{1}^T \mathbf{q} = d_\lambda + s_\lambda Q \qquad [14]$$

and the dipole polarizability remains invariant, which are exactly the translational symmetries required by classical electrostatics.[25,26] Importantly, the required physical symmetries are obtained without any specific choice of coordinate origin.

These solutions also obey the correct rotational symmetries. Under the global coordinate rotation $\vec{\mathbf{R}}_\lambda \mapsto U_{\lambda\rho}\vec{\mathbf{R}}_\rho$ as described by some rotation matrix $U \in SO(3)$, the dipole moment transforms as

$$d_\lambda \mapsto -U_{\lambda\rho}\mathbf{R}_\rho^T \boldsymbol{\eta}^{-1}(\boldsymbol{\chi} + \mu\mathbf{1}) = U_{\lambda\rho}d_\rho \qquad [15]$$

and the dipole polarizability transforms as

$$\alpha_{\lambda\rho} = -U_{\rho\sigma}\mathbf{R}_\sigma^T\boldsymbol{\eta}^{-1}U_{\lambda\tau}\mathbf{R}_\tau - \frac{\left(\mathbf{1}^T\boldsymbol{\eta}^{-1}U_{\rho\sigma}\mathbf{R}_\lambda\right)\left(\mathbf{1}^T\boldsymbol{\eta}^{-1}U_{\lambda\tau}\mathbf{R}_\tau\right)}{\mathbf{1}^T\boldsymbol{\eta}^{-1}\mathbf{1}} = U_{\rho\sigma}U_{\lambda\tau}\alpha_{\sigma\tau} \qquad [16]$$

which are exactly the transformational properties required of first– and second–rank tensors respectively.[26] Unlike the previous analysis for the translational symmetries, the rotational properties remain unchanged when the charge constraint terms are discarded.

### 4   Conclusion

We have investigated the separate effects of charge conservation from the much better appreciated effect of electronegativity imbalances in the electronegativity equalization process. The charge distribution of Eq. [7] calculated analytically using the method of Lagrange multipliers and Gaussian elimination show the contribution of two distinct terms that represent each effect separately. These distinct terms also give rise to

well–distinguished terms in the formulae for the dipole moment in Eq. [8] and dipole polarizability in Eq. [13], which automatically obey the requisite translational and rotational symmetries without the need for special choices of coordinate origin. In particular, we have seen that under spatial translations, the charge conservation terms give rise to counterterms that exactly cancel the unphysical terms arising from the electronegativity imbalance terms, which by themselves result in violation of the physically correct translational symmetries.

## 5 Acknowledgment

This work was supported by DOE under Grant No. DE-FG02-05ER46260.

## 6 References

[1] W. J. Mortier, S. K. Ghosh, and S. Shankar, J. Am. Chem. Soc. **108** (15), 4315 (1986).
[2] A. K. Rappé and W. A. Goddard, III, J. Phys. Chem. **95**, 3358 (1991).
[3] S. W. Rick and S. J. Stuart, in *Rev. Comp. Chem.*, edited by K. B. Lipkowitz and D. B. Boyd (Wiley, New York, 2002), Vol. 18, pp. 89.
[4] S. M. Valone and S. R. Atlas, Phil. Mag. **86**, 2683 (2006).
[5] S. W. Rick, S. J. Stuart, and B. J. Berne, J. Chem. Phys. **101**, 6141 (1994).
[6] D. M. York and W. Yang, J. Chem. Phys. **104**, 159 (1996).
[7] R. Chelli, P. Procacci, R. Righini, and S. Califano, J. Chem. Phys. **111**, 8569 (1999).
[8] S. W. Rick, J. Chem. Phys. **114** (5), 2276 (2001).
[9] S. J. Stuart and B. J. Berne, J. Phys. Chem. **100** (29), 11934 (1996).
[10] S. W. Rick and B. J. Berne, J. Am. Chem. Soc. **118** (3), 672 (1996).
[11] W. J. Mortier, K. Vangenechten, and J. Gasteiger, J. Am. Chem. Soc. **107** (4), 829 (1985).
[12] R. T. Sanderson, Science **114**, 670 (1951).
[13] F. H. Streitz and J. W. Mintmire, Phys. Rev. B **50** (16), 11996 (1994).
[14] E. P. Gyftopoulos and G. N. Hatsopoulos, Proc. Natl. Acad. Sci. **60**, 786 (1968).
[15] R. S. Mulliken, J. Chem. Phys. **2**, 782 (1934).
[16] R. G. Parr, R. A. Donnelly, M. Levy, and W. E. Palke, J. Chem. Phys. **68**, 3801 (1978).
[17] R. G. Parr and R. G. Pearson, J. Am. Chem. Soc. **105**, 7512 (1983).
[18] J. Chen, D. Hundertmark, and T. J. Martínez, J. Chem. Phys. **129**, 214113 (2008).
[19] R. P. Iczkowski and J. L. Margrave, J. Am. Chem. Soc. **83** (17), 3547 (1961).

20   J. Cioslowski and B. B. Stefanov, J. Chem. Phys. **99** (7), 5151 (1993).
21   R. F. Nalewajski, J. Phys. Chem. **89**, 2831 (1985).
22   S. M. Valone and S. R. Atlas, Phys. Rev. Lett. **97**, 256402 (2006).
23   J. Nocedal and S. J. Wright, *Numerical Optimization*, 2 ed. (Springer, New York, 2006).
24   M. Benzi, G. H. Golub, and J. Liesen, in *Acta Numerica*, edited by A. Iserles (Cambridge, Cambridge, 2005), Vol. 14, pp. 1.
25   J. D. Jackson, *Classical Electrodynamics*, 3 ed. (John Wiley and Sons, 1999).
26   J. Applequist, Chem. Phys. **85** (2), 279 (1984).